\documentclass[final,5p,sort&compress]{elsarticle}

\usepackage{cleveref}
\usepackage{siunitx}
\usepackage{tabularx}

\begin{document} 

\sloppy

\begin{frontmatter}

\title{Pitch Controls the Flexibility of Helical Ribbons}
\author[1]{Lucas Prévost}
\author[1]{Anke Lindner}
\author[1]{Olivia du Roure}
\address[1]{Laboratoire de Physique et Mécanique des Milieux Hétérogènes (PMMH), UMR7636 CNRS, ESPCI Paris, PSL Research University, Sorbonne Université, Université Paris Cité, 75005 Paris, France}

\begin{abstract}
Helical objects are often implemented in electronic or mechanical micro-systems, requiring a precise understanding of their mechanical properties. 
While helices formed by cylindrical filaments have been intensely investigated, little is known about the role of the cross-section of the filament at the basis of the helical shape.  
We study experimentally the force-extension response of micro-helices fabricated from ultra-thin PMMA ribbons. 
Leveraging newly achieved control on the helix geometry, the influence of the helical pitch is quantified and a significant stiffening of the helical ribbons with increasing pitch is highlighted.
Two phenomena are identified: a mechanical transition from a regime dominated by twisting of the ribbon at small pitch to a bending-dominated regime at high pitch and a purely geometrical effect, specific to helical ribbons. 
Excellent agreement is found with a previously established analytical model of inextensible elastic strips. 
\end{abstract}


\end{frontmatter}

\section{Introduction}
Helices display very interesting mechanical properties. 
They are typically formed by slender filaments rolled into a helical shape, with the radius of the helix being much larger compared to the filament thickness. 
Thanks to their tightly packed geometry, they store a high amount of elastic energy in a small volume. 
Due to the scale separation between the helix radius and the filament thickness, high stretch ratios of the overall geometry can be achieved with comparatively very little material strain. 
Helices are common in nature, spanning several orders of magnitude in length, including double-stranded DNA, $\alpha$-helix in proteins, bacterial flagella \cite{lauga2009hydrodynamics}, cholesteric crystals \cite{zastavker1999self}, or plant tendrils \cite{gerbode2012cucumber}.
Synthetic helices are widespread at the macro-scale: helical springs can absorb shocks, measure forces, or serve as actuators. 
In recent years, considerable efforts have been dedicated to implement such structures at sub-millimetric scales and integrate them within electronic or mechanical micro-systems \cite{huang2015helices,ren2014review,liu2014helical,wan2018helical}.
In regards to these applications, it is critical to accurately characterize the mechanical properties of helices.

Many studies have investigated theoretically, numerically, or experimentally the mechanics of helices and in particular the force-extension behavior \cite{love2013treatise,smith2001tension,pham2014stretching,bell2006fabrication,chen2003mechanics,gao2006superelasticity,khaykovich2009thickness,wada2007stretching}.
But, despite some natural and many synthetic helices being constituted of ribbon-like filaments \cite{zastavker1999self,pham2013highly,grutzmacher2008ultra,zhang2009artificial,bell2006fabrication,li2012superelastic,zhang2017dynamic}, studies rarely investigated the specific case of helical ribbons and generally referred to cylindrical filaments, using classical rod models.
Recent works have highlighted that the mechanical properties of a ribbon, \textit{i.e.} a filament of highly anistropic cross-section, are qualitatively and quantitatively different to that of a rod, \textit{i.e.} a filament of istropic or near-isotropic cross-section (\textit{e.g.} circle, square), and thus requires specific investigation  \cite{audoly2016buckling,kumar2020investigation}.
Pham et al. \cite{pham2014stretching,pham2015deformation} probed experimentally the mechanical response of tightly coiled micron-sized helical ribbons of various materials and the work of Smith \textit{et. al} \cite{smith2001tension} has measured experimentally the spring constant of cholesteric helical ribbons. 
However, only the analytical and numerical work of Starostin et al. \cite{starostin2008tension} has specifically investigated the mechanical properties of helical ribbons and has systematically addressed the influence of the pitch on their mechanical response but experimental validation is still lacking.

We have recently developed a fabrication method for highly flexible micron-sized helical ribbons with full control of the geometrical parameters (helical radius, total length \emph{and} pitch) \cite{prevost2022shaping}.
In this work, we leverage this unprecedented control to study experimentally the influence of the pitch on the force-extension behavior of helical ribbons. 
To do so we clamp the helices between a micro-capillary and a cantilever. 
With the help of a micromanipulator the capillary is moved successively, imposing a well controlled extension of the micro-helix, whereas the deformation of the cantilever is used to detect the applied force with a nano-Newton resolution.  
We report a significant stiffening of helical ribbons as the pitch is increased (going from closed-coiled to open-coiled).
Our experimental results are compared to the analytical model of Starostin et al. \cite{starostin2008tension}, developed specifically for helical ribbons, finding excellent agreement. 

\section{Experimental Methods}

\subsection{General Principles}

\begin{figure}[b!]
    \centering
    \includegraphics[width = \linewidth]{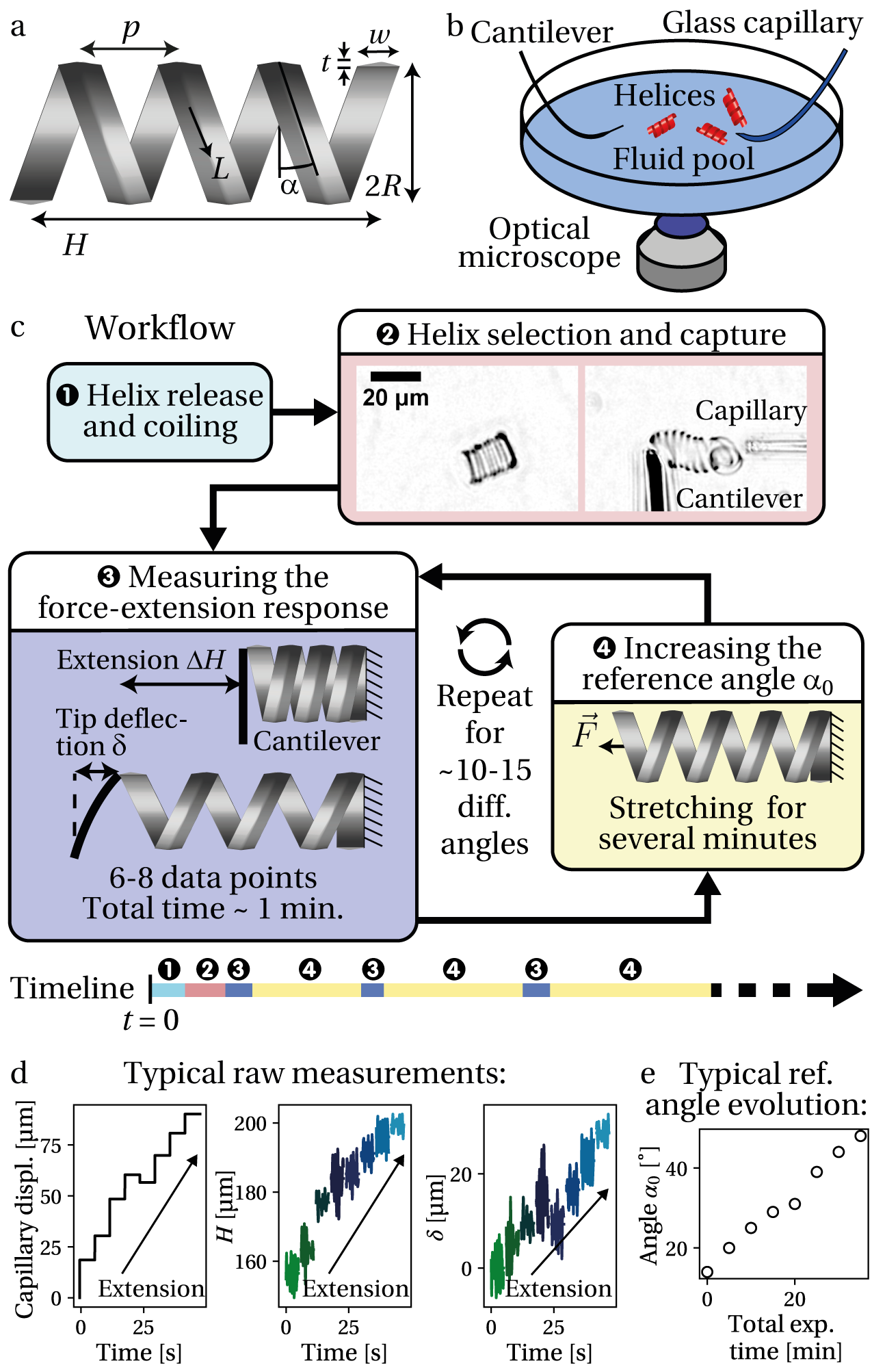}
    \caption{a. Helical ribbon geometry and relevant geometrical parameters. Typical values range within $R = \SIrange{1}{100}{\micro \meter}$, $L = \SIrange{100}{3000}{\micro \meter}$, $t = \SIrange{5}{50}{\nano \meter}$ and $w = \SIrange{0.5}{5}{\micro \meter}$ b. Schematic of the experimental set-up. c. Experimental workflow for the mechanical characterization of helical ribbons. Samples are immersed in water: ribbons lift-off following the dissolution of the sacrificial layer and self-coil into a tight helical shape. A helix is selected and captured by the open glass capillary. Contact is made between the tip of the cantilever and the other helix end. The two next steps are repeated 10 to 15 times: firstly establishing the force-extension curve for a given helix geometry and secondly increasing the pitch angle using the stretching treatment method. The timeline illustrates the succession of these different steps. d. Typical raw measurements obtained when establishing a force-extension curve: the capillary is successively displaced (left plot), which extends the helix. The extension is recorded by measuring the axial length $H$ (middle plot) and the force is obtained from the cantilever tip displacement $\delta$ (right plot). e. Typical evolution of the reference pitch angle $\alpha_0$ of a given helical ribbon over a full experiment.}
    \label{fig:flow}
\end{figure}

The helical ribbons are produced using a two-step fabrication method, described in detail in a previous publication \cite{prevost2022shaping}.
The first step relies on the spontaneous formation of highly flexible helical ribbons driven by surface tension \cite{pham2013highly}. 
Nanometer-thick ribbons are prepared on a flat sacrificial layer through an evaporative assembly method \cite{kim2010nanoparticle}.
As a result of the fabrication process, the ribbons display a near-triangular cross-section of width $w$ (typically $\SIrange{0.5}{5}{\micro \meter}$) and thickness $t$ (typically $\SIrange{5}{50}{\nano \meter}$), such that the aspect ratio of the cross-section verifies $t/w \ll 1$.
Upon release of the ribbons into a liquid, they quickly form a tightly coiled helical geometry.
The helical ribbon geometry is shown in \cref{fig:flow}.a, along with the relevant geometrical parameters. 
The pitch angle is defined such that $\tan \alpha = p / 2\pi R$, and is found to be close to zero after the initial helix formation \cite{pham2013highly}. 
The total filament length $L$ is directly controlled during the fabrication process and can be further tuned by cutting the ribbons prior to release in liquid. 
The helical radius $R$ is determined by a balance between surface tension and elasticity and scales as $R \sim E t^2/\gamma$, with $E$ being the Young's modulus and $\gamma$ the surface tension \cite{pham2013highly}. 
This affords control over the radius through modification of the ribbon thickness, either during ribbon fabrication or by their subsequent etching \cite{choudhary2019controlled}.

Control of the pitch angle $\alpha$ is achieved in a second step by leveraging the creep properties of the material to shape the helical ribbon into the desired geometry. 
Sizable creep effects are not expected in bulk for the materials we use but are enabled by the nanoscale confinement in the ribbon \cite{paeng2011molecular,bansal2005quantitative}.
A stress is applied to the material by extending the helix for a long period of time, typically several minutes, after which the helix is allowed to relax. 
During this application of uniform stress, achieved by end-loading the helix, an irreversible \emph{uniform} increase in the pitch angle is observed. 
This process, executed \textit{in-situ}, is termed the 'stretching treatment' \cite{prevost2022shaping}.

We take advantage of this \textit{in-situ} procedure to investigate the influence of the pitch angle on the force-extension response, keeping all other parameters fixed.
In the work presented here, we alternate, for a given helical ribbon, a mechanical characterization and a stretching treatment, thus obtaining the evolution of the helix mechanical properties as the pitch angle is increased. 
By working with a single ribbon throughout an experiment, the ribbon thickness $t$ and width $w$, which are difficult to accurately measure or control due to the small size, will remain unchanged, as well as the total filament length $L$.
The radius $R$ may vary slightly as a by-product of stretching but as the influence of the radius on the pulling force is well established ($F \propto 1/R^2$) \cite{pham2013highly,smith2001tension,starostin2008tension}, these variations are easily corrected.

\subsection{Experimental Set-up}
The ribbons are prepared on a water-soluble sacrificial layer, following the method established by Lee et al. \cite{lee2013macroscopic}, then released into a pool of water (see \cref{fig:flow}.b for the full experimental set-up).
The ribbons are made of poly(methyl methacrylate) (PMMA, $M_\text{w} = \SI{120d3}{}$) prepared on a sacrificial layer of poly(acrylic acid) (PAA, $M_\text{w} = 1800$).
An open glass capillary tube is connected to a syringe pump and fixed to a micromanipulator to allow capture and release of the helices by pumping or expelling liquid.
A carbon fiber cantilever is similarly fixed to a second micromanipulator to afford force measurements. 
The experimental set-up is mounted on an inverted optical microscope connected to a numerical camera.

Prior to experiments, the bending modulus of the carbon fiber cantilever is calibrated at $B_\text{cant} = \SI{1.55 \pm 0.02 d-11}{\pascal \meter \tothe{4}}$.
When the cantilever is submitted to a force $F$ perpendicular to its direction, the tip deflection is given by $\delta  = FL_\text{cant}^3/3B_\text{cant}$.
With typical cantilever length $L_\text{cant} \sim \SI{1}{\centi \meter}$ and an optical resolution of a few microns, sub-nanonewton forces can be measured. 

\subsection{Experimental Workflow}

The experimental workflow is illustrated in \cref{fig:flow}.c.
After release and coiling of the helices, a helix is selected and one of its ends is caught and clamped by the open glass capillary.
The cantilever is approached and contact is made between the cantilever tip and the other end of the helix.
Non-specific contact forces afford strong adhesion between the cantilever and the helix.
Two steps are then repeated until the end of the experiment: first measuring the force-extension response at a given geometry and then increasing the pitch angle using the stretching treatment.
Both steps are further detailed in the following.

To measure the force-extension response, a series of increasing extension steps is imposed to the helix by successively displacing the capillary.
The capillary is quickly displaced from one position to another (typically 6 to 8 positions, with typical displacement $\sim \SIrange{10}{20}{\micro \meter}$ ) and is held still for a few seconds at each position (typically 5 seconds). 
At each position and for the few seconds the capillary is held still, the geometry of the deformed helix and the cantilever tip position are recorded.
From the experimental images and knowing the reference position of the cantilever tip (\textit{i.e.} when no force is applied), the cantilever tip displacement $\delta$ and the helix axial length $H$ are extracted. 

We show in \cref{fig:flow}.d a typical measurement for $\delta$ and $H$ obtained when measuring a force-extension response. 
As expected, as the axial length is further increased, the cantilever tip is further deflected: the force applied to stretch the helix is recorded by the cantilever. 
We observe that noise can be significant, which probably originates from ambient flow in the fluid pool combined with the very long length of the cantilever fiber, thus creating fluctuations in the cantilever tip position. 
As the helix is held between the cantilever tip and the capillary, these fluctuations also impact the measurement of the helix axial length.
To mitigate this noise, the cantilever tip deflection $\delta$ and the helix axial length $H$ are averaged over the 5 seconds recordings.
The tension force $F$ is then computed from the cantilever length and bending modulus: $F = 3 \delta B_\text{cant}/L_\text{cant}^3$.

Before and after each force-extension measurement, the helix is released from the open glass capillary and its resting axial length is measured.
Due to the creep properties of the material (leveraged to control the pitch angle) an increase in the resting axial length is observed.
We mitigate these effects by keeping the experiment duration as short as possible: typically $\sim \SIrange{30}{40}{\second}$ for a given force-extension curve (the typical timescale of creep deformations is several minutes). 
As a result, the difference between the two measurements of the resting axial length is always below $\sim \SI{2}{\percent}$ of the total length $L$, which corresponds to a variation in pitch angle of less than $\SI{1}{\degree}$, which we neglect. 
The helix reference axial length $H_0$, taken as the average over the two measurements, is thus considered constant for a given force-extension curve. 
The reference pitch angle is simply obtained from the geometrical relation $H_0/L = \sin \alpha_0$.

We then apply the stretching treatment to increase the reference pitch angle $\alpha_0$.
The helix is clamped at its two ends (one by the cantilever and one by the capillary), an axial extension (typically $\sim \SIrange{50}{100}{\micro \meter})$ is imposed for several minutes by displacing the capillary, and the helix is finally let to relax. 
This treatment results in an irreversible increase in the reference pitch angle. 
We show in \cref{fig:flow}.e the typical evolution of the reference pitch angle over a full experiment.

A new force-extension curve is then measured for this new reference geometry and so on.
At the end of the experiment, the helix is completely stretched to measure the total filament length $L$. 
This experimental workflow yields a series of force-extension curves, each corresponding to an increasing reference pitch angle $\alpha_0$.

\section{Experimental Results}
\subsection{Force-Extension Curves}

\begin{figure}[t!]
    \centering
    \includegraphics[width = \linewidth]{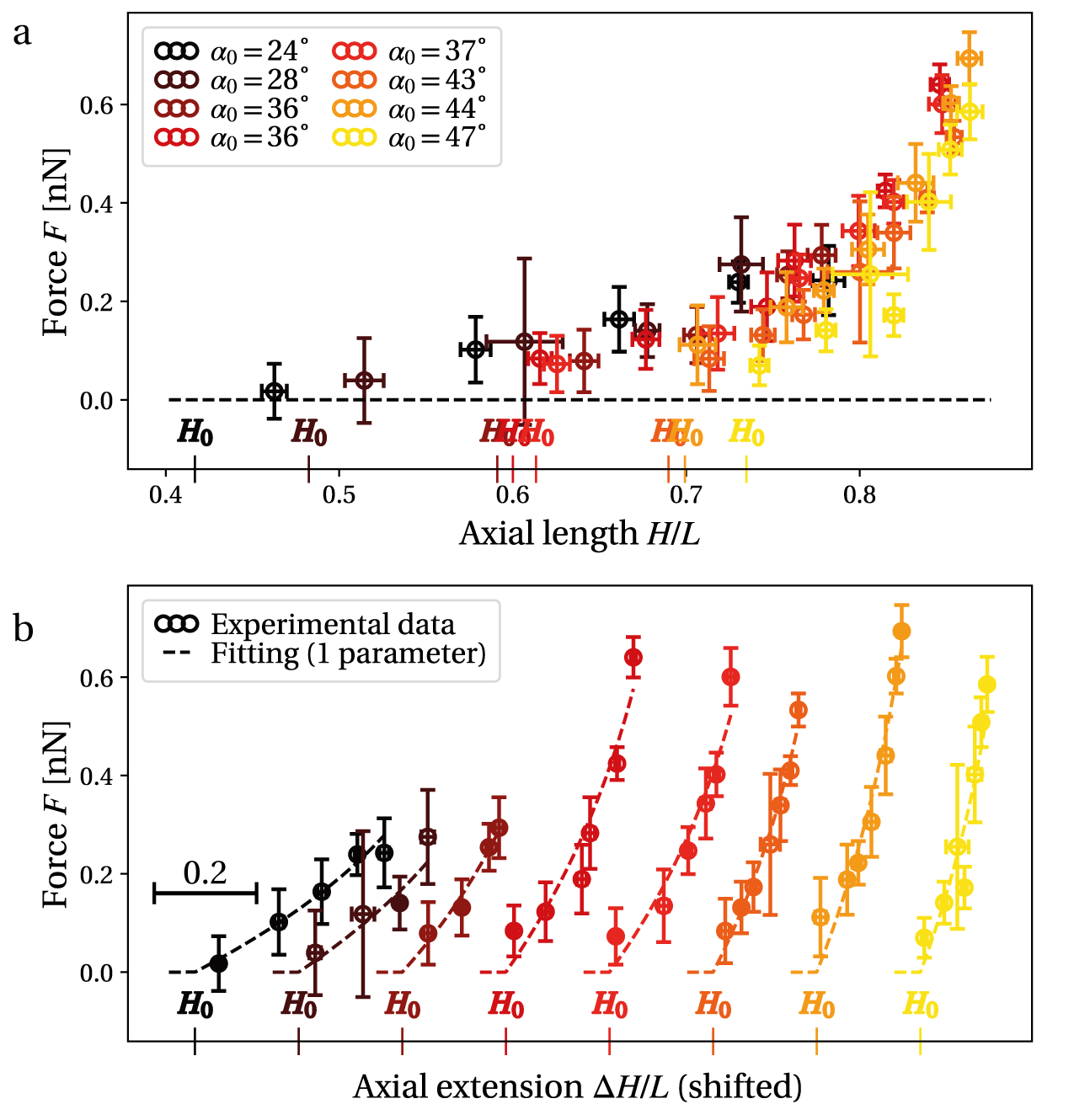}
    \caption{Experimental results for one PMMA helical ribbon (with total length $L = \SI{480}{\micro \meter}$ and radius $R = \SI{5.0}{\micro \meter}$) a. Pulling force $F$ as a function of rescaled axial length $H/L$ as the geometry is varied. Each curve is associated to a different reference axial length $H_0$, shown at the bottom of the plot, and thus to a different reference pitch angle $\alpha_0$. The error bars are the standard deviations calculated over the 5 seconds recordings. b. Successive force-extension curves fitted with the analytical model proposed by Starostin et al. \cite{starostin2008tension}. The curves are shifted by an arbitrary amount to better distinguish between the different curves.}
    \label{fig:results}
\end{figure}

We show in \cref{fig:results}.a the measured force $F$, computed from the cantilever tip deflection $\delta$, as a function of the rescaled axial length $H/L$, for a given helix as the helix geometry is varied. 
Each curve on the plot corresponds to a different reference pitch angle $\alpha_0$.
The corresponding reference axial length $H_0$ for each geometry is represented at the bottom of the plot. 

We observe that the helical ribbon stiffens as the reference pitch angle is increased. 
Indeed, for the light-colored curves, which correspond to higher reference pitch angles, a comparatively higher force is necessary to extend the helix than for the dark-colored ones.
The same observation is confirmed for other helical ribbons, as shown in the SI. 
To explain this stiffening, the relation between the helix geometry and the force-extension response must be understood. 

As shown in \cref{fig:results}.a, the imposed extension can be significant, the axial length reaching up to $\sim \SI{95}{\percent}$ of the total length $L$ \textit{i.e.} almost to full extension.
In this high stretch regime, close to full helix extension (corresponding to $H/L = 1$), the force-extension relationship is not linear. 
Due to the specific geometrical properties of the helical object, such significant global deformation is obtained while remaining in the material elastic regime \cite{prevost2022shaping}. 
The non-linearity of the force-extension relationship thus does not originate from the material itself but from the helical geometry. 
The analytical work of Starostin et al. \cite{starostin2008tension} has examined in the general case the elastic deformation of a helical ribbon submitted to an axial end-loading. 
Under the assumption that the deformed helix remains a uniform helix (uniform radius and angle), they obtained the following expression for the force-extension relationship
\begin{equation} \label{eq:fit}
    F = \frac{C}{R^2} \frac{\left( \cos \alpha_0 \cos \alpha + \frac{C}{B} \sin \alpha_0 \sin \alpha \right)}{\left( \cos^2 \alpha + \frac{C}{B} \sin^2 \alpha \right)^2} \frac{\sin (\alpha - \alpha_0)}{\cos \alpha}
\end{equation}
with $\alpha = \arcsin H/L$ pitch angle of the deformed helix (supposedly uniform along the filament length), which thus describes the deformation of the helix. 
Because of the multiplicative term $1/\cos \alpha$, the force diverges as the helix is fully extended (corresponding to $\alpha \rightarrow \SI{90}{\degree}$), which describes the helix finite extensibility. 

For a flat triangular cross-section, we have $B = E w t^3/36$ and $C = \mu w t^3 / 12$ \cite{gloumakoff1964} with $\mu$ shear modulus and hence $C/B = 3/2(1+\nu)$ \textit{i.e.} independent of the ribbon thickness and width. 
The Poisson's ratio $\nu$ for bulk PMMA is usually estimated within the range $\nu = \SIrange{0.35}{0.4}{}$ \cite{gilmour1979elastic}. 
We consider that the Poisson's ratio is not affected by the strong material confinement due to the ribbon vanishing thickness and we take $\nu = 0.375$.

In theory, the twisting modulus $C$ could be measured prior to experiments by measuring the ribbon thickness and width.
But in practice, accurate measurement of the ribbon nanoscale thickness prior to experiments is very difficult.
Furthermore, determining the material shear modulus $\mu$ would require further investigation as the bulk value is most likely not relevant: nanoscale thicknesses have been shown to influence the mechanical properties of materials \cite{stafford2006elastic,torres2009elastic,bay2020mechanical}.
We thus use $C$ as a fitting parameter.

We show in \cref{fig:results}.b the force-extension curves (same data as in \cref{fig:results}.a) fitted using \cref{eq:fit}.
The successive curves are fitted independently, each fitting yielding an estimate of the twisting modulus $C$.
The curves are shifted by an arbitrary amount to better distinguish between the different curves.
For all reference pitch angles, the proposed expression accurately fits the experimental data and describes the non-linearity of the force-extension curves.
The experimental protocol is repeated for three other helical ribbons (4 in total): we present in the SI the fitted force-extension curves for the other helical ribbons.
Excellent agreement with the theoretical prediction is obtained for all four, using in all cases $C$ as the single fitting parameter.
Our experimental results thus provide strong validation for the analytical relationship obtained by Starostin et al. \cite{starostin2008tension}.
To further verify the consistency of these results, we examine the obtained values of $C$.

\subsection{Measurements of the Twisting Modulus}

\begin{figure}[t!]
    \centering
    \includegraphics[width = \linewidth]{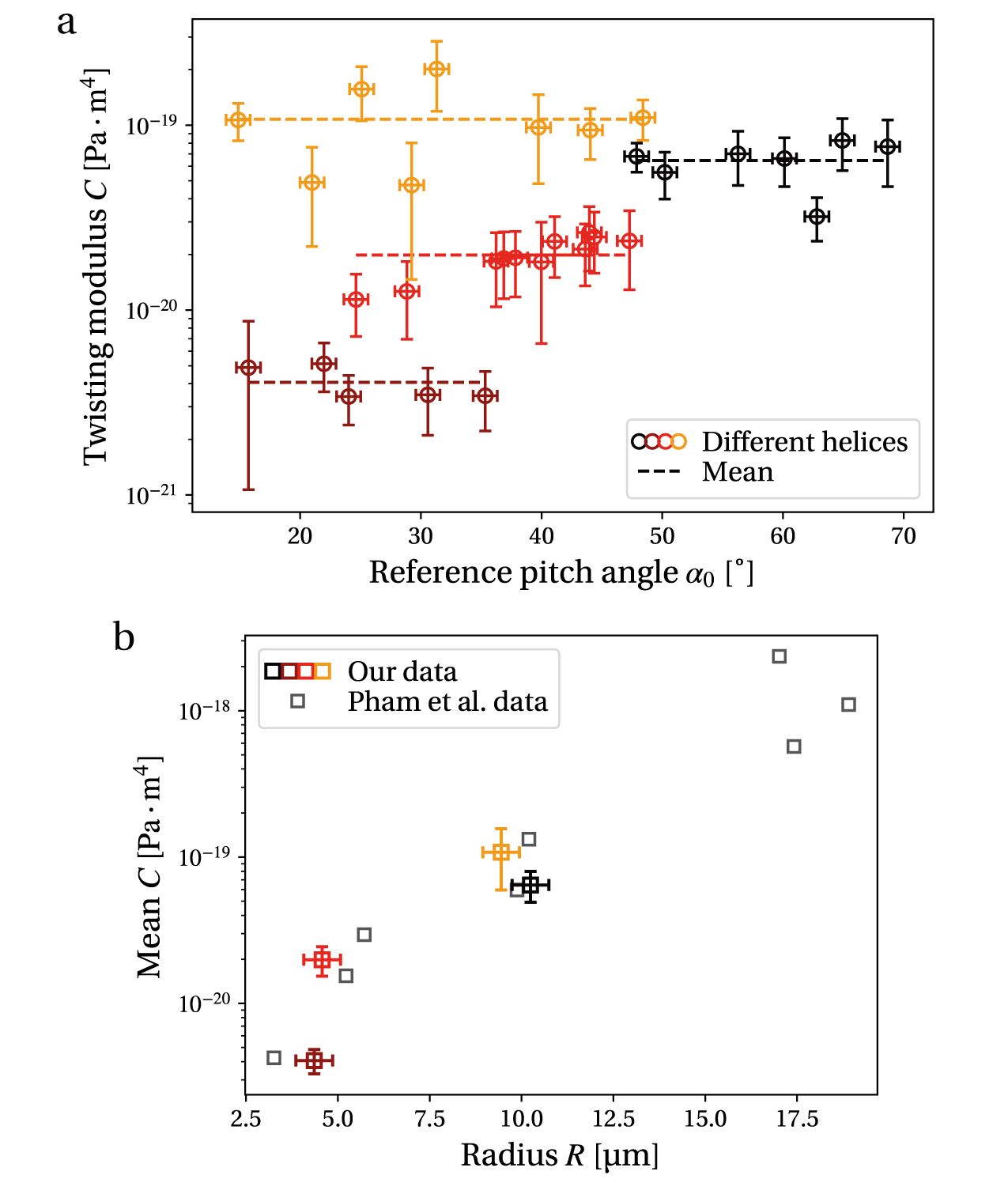}
    \caption{a Estimate of the twisting modulus $C$ obtained from fitting the force-extension curves as a function of reference pitch angle for 4 different PMMA helical ribbons. The error bars represent the numerical uncertainties from the fitting. Data shown in \cref{fig:results} correspond to the red squares. b. Twisting modulus $C$ of each helical ribbon, averaged over all reference pitch angles, as a function of initial helical radius $R$. Colored dots show our experimental results while gray dots show the results of Pham et al. (also obtained for PMMA helical ribbons) \cite{pham2015deformation}.}
    \label{fig:modulus}
\end{figure}

At each reference pitch angle $\alpha_0$, the fitting yields one estimate of the ribbon twisting modulus $C$.
Naturally, this parameter is expected to remain constant for a given helix throughout the experiment. 
We show in \cref{fig:modulus}.a the successive measurements for $C$ as the reference pitch angle $\alpha_0$ is increased, for the four different helical ribbons. 
In all four cases and within the experimental errors, no change is observed for the twisting modulus. 
Overall, the reference pitch angle spans a wide range $\sim \SIrange{10}{70}{\degree}$.
These results further validate the model proposed by Starostin et al. \cite{starostin2008tension}: the stiffening observed as the reference pitch angle increases is accurately captured by the proposed expression. 

The work of Pham et al. \cite{pham2015deformation} previously studied the mechanical response of PMMA helical ribbons, with the same experimental system.
The twisting modulus $C$ was measured for several PMMA helical ribbons of various radii $R$.
These results were obtained for helices in their initial configuration after coiling \textit{i.e.} with vanishing reference pitch angles $\alpha_0$.
At small reference angle ($\alpha_0 \ll 1$) and within the small deformation regime ($\alpha - \alpha_0 \ll 1$),
the force-extension relationship \cref{eq:fit} simplifies to $F = (C/R^2) * (\Delta H/L)$.
Following the common hypothesis $B=C$ \cite{kim2005deformation}, Pham et al. used a different relation $F = (B/R^2) * (\Delta H/L)$.
Their results are thus presented as measurements of the bending modulus $B$ while they actually measured the twisting modulus $C$.
We reproduce in \cref{fig:modulus}.b the results of Pham et al. along with our results, finding excellent consistency, further supporting the experimental validation of \cref{eq:fit}.

\section{Discussion}
\begin{figure}[b!]
    \centering
    \includegraphics[width = \linewidth]{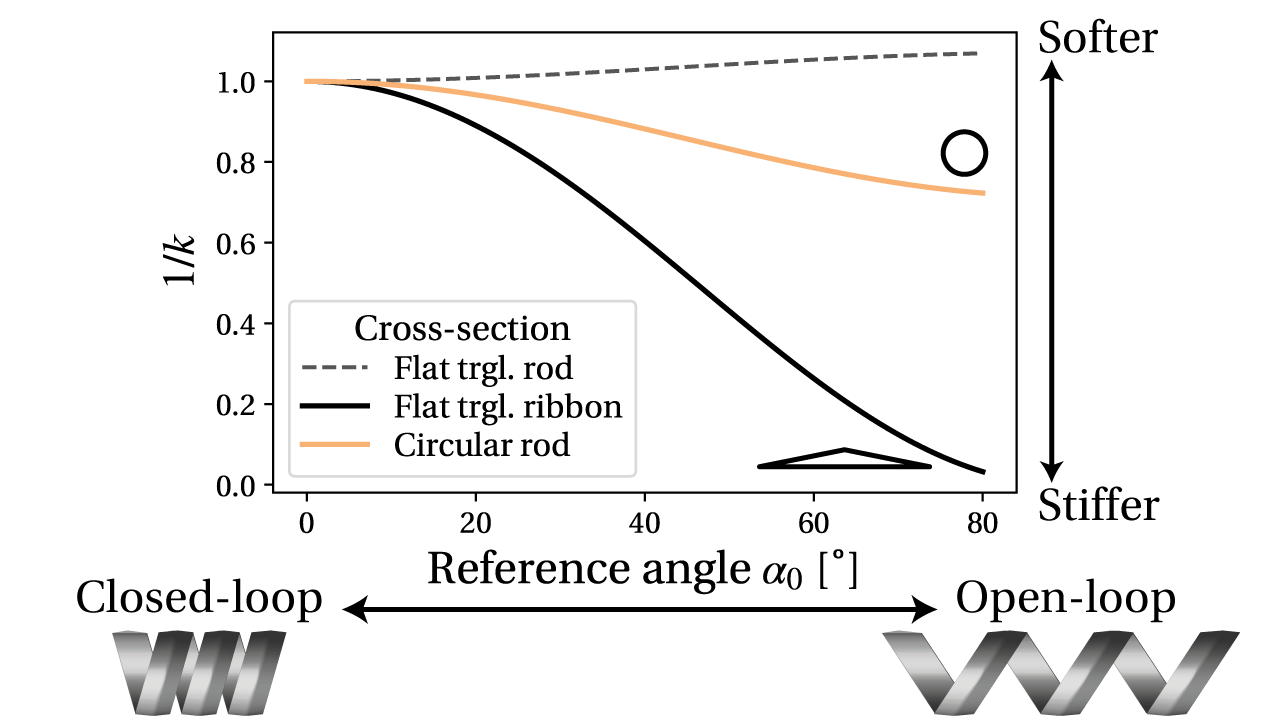}
    \caption{
    Helix linear stiffness for a flat triangular filament and for a circular filament, represented by the inverse of the normalized spring constant $1/k$ as a function of reference pitch angle $\alpha_0$. The dashed line represents the fictive case of a flat triangular rod \textit{i.e.} without the stiffening term. The Poisson's ratio was taken as $\nu = 0.4$ (since for most materials $\nu \sim \SIrange{0.3}{0.5}{}$).
    }
    \label{fig:synthesis}
\end{figure}

In order to explain the stiffening of the helical ribbons, it is beneficial to write the linear limit of the force-extension relationship \textit{i.e.} when $\alpha - \alpha_0 \ll 1$.
Within this limit, the expression reads
\begin{equation}
    F = \frac{1}{R^2} \frac{1}{\left( \frac{1}{C} \cos^2 \alpha_0 + \frac{1}{B}\sin^2 \alpha_0 \right) \cos^2 \alpha_0 } \frac{\Delta H}{L}
    \label{eq:ribbon}
\end{equation}
In the case of filaments with isotropic or near-isotropic cross-sections \textit{i.e.} rods, the force-extension relationship has been found to be \cite{love2013treatise}
\begin{equation}
     F_\text{iso} = \frac{1}{R^2} \frac{1}{\left( \frac{1}{C} \cos^2 \alpha_0 + \frac{1}{B}\sin^2 \alpha_0 \right)} \frac{\Delta H}{L}
     \label{eq:rod}
\end{equation}
The only difference between the ribbon case and the rod case is thus the multiplicative term $1/\cos^2 \alpha_0$.
The common term $ 1/\left(\frac{1}{C} \cos^2 \alpha_0 + \frac{1}{B}\sin^2 \alpha_0 \right)$ denotes a mechanical transition from a regime dominated by twisting of the ribbon ($C$ being the relevant modulus) at small $\alpha_0$ (\textit{i.e.} closed-loop helices) to a bending-dominated regime ($B$ being the relevant modulus) at high $\alpha_0$ (\textit{i.e.} open-loop helices). 
In the case of a circular cross-section, we have $C/B  = 1/(1+\nu)$ and hence $C/B \sim 0.7$ for most materials (since usually $\nu \sim \SIrange{0.3}{0.5}{}$): this term drives a weak stiffening with increasing $\alpha_0$.
In our experimental conditions we have $C/B = 3/2(1+\nu) \sim 1.1$: this term alone would thus drive a very weak \emph{softening} as $\alpha_0$ increases.
Therefore, the significant stiffening observed for helical ribbons stems from the second term $1/\cos^2 \alpha_0$, which dominates the other and is not found in the isotropic case.
We conclude that this strong stiffening behavior is specific to helical ribbons. 
\Cref{fig:synthesis} synthesizes these two different behaviors by representing the inverse of the normalized spring constant $1/k$, with $k$ defined such that $\displaystyle F = k(\alpha_0) \frac{C}{R^2} \frac{\Delta H}{L}$, as a function of reference pitch angle $\alpha_0$ for a flat triangular filament and for a circular filament.
The dashed line represents \cref{eq:rod} (\textit{i.e.} without the $1/\cos^2 \alpha_0$ term) for a flat triangular filament ($C/B \sim 1.1$) to illustrate the weak softening induced by the twisting-to-bending mechanical transition in our experimental conditions. 
We call this fictive case the 'flat triangular rod'. 

Our investigation provides strong experimental validation to the model proposed by Starostin et al. \cite{starostin2008tension}.
This validation comes despite several differences between the assumptions of Starostin et al. and our experimental conditions.
They assumed a force-controlled stretching while in our experiments the stretching is displacement-controlled (the applied force being measured through the cantilever). 
This entails differences in the boundary conditions: in our conditions both helix ends are clamped and thus the applied force is not purely axial, a tangential component may also exist. 
This tangential component is not recorded, as it is applied parallel to the cantilever axis.
Thus, the very good agreement between the predictions of Starostin et al. and our experimental results suggests that boundary conditions have only a weak impact on the helix global mechanical response. 
We strongly suspect that, similarly to what was found in the case of rods \cite{kim2005deformation}, this is due to the typical filament length being much larger than the typical helical radius.
In the case of shorter helices, boundary conditions may have a more pronounced influence. 
Finally, Starostin et al. based their analytical model on the case of flat rectangular filaments and validated their model against numerical simulations of such helices, while in our experimental conditions the cross-section is near-triangular. 
Our results thus suggests validity of the model for ribbons of all cross-sectional shapes: flat triangle or flat ellipse for example.

\section{Conclusion}

In this work, we have characterized experimentally the mechanical properties of micron-sized helical ribbons by measuring the force-extension relationship.
An axial extension was imposed, and the corresponding force was measured using a cantilever beam.
Benefiting from newly achieved control on the pitch angle, the influence of this parameter was investigated experimentally in detail.
At a given reference pitch angle, the force-extension response was found to be non-linear. 
As the reference pitch angle $\alpha_0$ was increased, a strong stiffening of the helical ribbons was observed. 
We found that both the non-linearity and the stiffening originate from the helical geometry and are accurately described by the analytical model proposed by Starostin et al. \cite{starostin2008tension}.
Our results thus provide strong experimental validation for the obtained force-extension relationship, given by \cref{eq:fit}. 
Comparison with a model previously established for filaments of isotropic cross-section \cite{love2013treatise} highlighted an effect common to rods and filaments: a twisting-to-bending transition as the pitch increases. 
But, while this effect stiffens helices formed by rods, it tends to soften helical ribbons.
Therefore, the observed strong stiffening effect is specific to helical ribbons. 
Modifying the filament cross-section can thus be used in the future as a design principle to fabricate helices with specific mechanical properties. 
We hope that these findings stimulate further studies of the intermediate case \textit{i.e.} filaments whose cross-section is neither isotropic nor flat as to clarify the origin of this stiffening behavior and to explain the crossover between the two models proposed. 

\section*{Acknowledgments}
We thank Prof. A. J. Crosby for stimulating discussions and a critical reading of the manuscript. 
We thank Dr. D. M. Barber for help with the helix fabrication.
AL and LP acknowledge funding by the European Research Council through a consolidator grant (ERC PaDyFlow 682367). 
This work received the support of Institut Pierre-Gilles de Gennes (Équipement d’Excellence, “Investissements d’Avenir”, Program ANR-10-EQPX-34).

\section*{Conflict of Interest}
The authors declare no conflict of interest.

\bibliography{bib_mech}

\begin{thebibliography}{10}
\expandafter\ifx\csname url\endcsname\relax
  \def\url#1{\texttt{#1}}\fi
\expandafter\ifx\csname urlprefix\endcsname\relax\def\urlprefix{URL }\fi
\expandafter\ifx\csname href\endcsname\relax
  \def\href#1#2{#2} \def\path#1{#1}\fi

\bibitem{lauga2009hydrodynamics}
E.~Lauga, T.~R. Powers, The hydrodynamics of swimming microorganisms, Reports
  on Progress in Physics 72~(9) (2009) 096601.

\bibitem{zastavker1999self}
Y.~V. Zastavker, N.~Asherie, A.~Lomakin, J.~Pande, J.~M. Donovan, J.~M. Schnur,
  G.~B. Benedek, Self-assembly of helical ribbons, Proceedings of the National
  Academy of Sciences 96~(14) (1999) 7883--7887.

\bibitem{gerbode2012cucumber}
S.~J. Gerbode, J.~R. Puzey, A.~G. McCormick, L.~Mahadevan, How the cucumber
  tendril coils and overwinds, Science 337~(6098) (2012) 1087--1091.

\bibitem{huang2015helices}
G.~Huang, Y.~Mei, Helices in micro-world: Materials, properties, and
  applications, Journal of Materiomics 1~(4) (2015) 296--306.

\bibitem{ren2014review}
Z.~Ren, P.-X. Gao, A review of helical nanostructures: growth theories,
  synthesis strategies and properties, Nanoscale 6~(16) (2014) 9366--9400.

\bibitem{liu2014helical}
L.~Liu, L.~Zhang, S.~M. Kim, S.~Park, Helical metallic micro-and
  nanostructures: fabrication and application, Nanoscale 6~(16) (2014)
  9355--9365.

\bibitem{wan2018helical}
G.~Wan, C.~Jin, I.~Trase, S.~Zhao, Z.~Chen, Helical structures mimicking chiral
  seedpod opening and tendril coiling, Sensors 18~(9) (2018) 2973.

\bibitem{love2013treatise}
A.~E.~H. Love, A treatise on the mathematical theory of elasticity, Cambridge
  university press, 1944.

\bibitem{smith2001tension}
B.~Smith, Y.~V. Zastavker, G.~B. Benedek, Tension-induced straightening
  transition of self-assembled helical ribbons, Physical review letters 87~(27)
  (2001) 278101.

\bibitem{pham2014stretching}
J.~T. Pham, J.~Lawrence, G.~M. Grason, T.~Emrick, A.~J. Crosby, Stretching of
  assembled nanoparticle helical springs, Physical Chemistry Chemical Physics
  16~(22) (2014) 10261--10266.

\bibitem{bell2006fabrication}
D.~J. Bell, L.~Dong, B.~J. Nelson, M.~Golling, L.~Zhang, D.~Gr{\"u}tzmacher,
  Fabrication and characterization of three-dimensional ingaas/gaas
  nanosprings, Nano Letters 6~(4) (2006) 725--729.

\bibitem{chen2003mechanics}
X.~Chen, S.~Zhang, D.~A. Dikin, W.~Ding, R.~S. Ruoff, L.~Pan, Y.~Nakayama,
  Mechanics of a carbon nanocoil, Nano Letters 3~(9) (2003) 1299--1304.

\bibitem{gao2006superelasticity}
P.~X. Gao, W.~Mai, Z.~L. Wang, Superelasticity and nanofracture mechanics of
  zno nanohelices, Nano letters 6~(11) (2006) 2536--2543.

\bibitem{khaykovich2009thickness}
B.~Khaykovich, N.~Kozlova, W.~Choi, A.~Lomakin, C.~Hossain, Y.~Sung, R.~R.
  Dasari, M.~S. Feld, G.~B. Benedek, Thickness--radius relationship and spring
  constants of cholesterol helical ribbons, Proceedings of the National Academy
  of Sciences 106~(37) (2009) 15663--15666.

\bibitem{wada2007stretching}
H.~Wada, R.~R. Netz, Stretching helical nano-springs at finite temperature, EPL
  (Europhysics Letters) 77~(6) (2007) 68001.

\bibitem{pham2013highly}
J.~T. Pham, J.~Lawrence, D.~Y. Lee, G.~M. Grason, T.~Emrick, A.~J. Crosby,
  Highly stretchable nanoparticle helices through geometric asymmetry and
  surface forces, Advanced Materials 25~(46) (2013) 6703--6708.

\bibitem{grutzmacher2008ultra}
D.~Gr{\"u}tzmacher, L.~Zhang, L.~Dong, D.~Bell, B.~Nelson, A.~Prinz, E.~Ruh,
  Ultra flexible sige/si/cr nanosprings, Microelectronics journal 39~(3-4)
  (2008) 478--481.

\bibitem{zhang2009artificial}
L.~Zhang, J.~J. Abbott, L.~Dong, B.~E. Kratochvil, D.~Bell, B.~J. Nelson,
  Artificial bacterial flagella: Fabrication and magnetic control, Applied
  Physics Letters 94~(6) (2009) 064107.

\bibitem{li2012superelastic}
W.~Li, G.~Huang, J.~Wang, Y.~Yu, X.~Wu, X.~Cui, Y.~Mei, Superelastic metal
  microsprings as fluidic sensors and actuators, Lab on a Chip 12~(13) (2012)
  2322--2328.

\bibitem{zhang2017dynamic}
H.~Zhang, A.~Mourran, M.~Moller, Dynamic switching of helical microgel ribbons,
  Nano letters 17~(3) (2017) 2010--2014.

\bibitem{audoly2016buckling}
B.~Audoly, K.~A. Seffen, Buckling of naturally curved elastic strips: the
  ribbon model makes a difference, in: The Mechanics of Ribbons and M{\"o}bius
  Bands, Springer, 2016, pp. 293--320.

\bibitem{kumar2020investigation}
A.~Kumar, P.~Handral, C.~D. Bhandari, A.~Karmakar, R.~Rangarajan, An
  investigation of models for elastic ribbons: Simulations \& experiments,
  Journal of the Mechanics and Physics of Solids 143 (2020) 104070.

\bibitem{pham2015deformation}
J.~T. Pham, A.~Morozov, A.~J. Crosby, A.~Lindner, O.~du~Roure, Deformation and
  shape of flexible, microscale helices in viscous flow, Physical Review E
  92~(1) (2015) 011004.

\bibitem{starostin2008tension}
E.~Starostin, G.~van~der Heijden, Tension-induced multistability in
  inextensible helical ribbons, Physical review letters 101~(8) (2008) 084301.

\bibitem{prevost2022shaping}
L.~Prévost, D.~M. Barber, M.~Daïeff, J.~T. Pham, A.~J. Crosby, T.~Emrick,
  O.~du~Roure, A.~Lindner, Shaping nanoscale ribbons into microhelices of
  controllable radius and pitch, ACS Nano 16~(7) (2022) 10581--10588.

\bibitem{kim2010nanoparticle}
H.~S. Kim, C.~H. Lee, P.~Sudeep, T.~Emrick, A.~J. Crosby, Nanoparticle stripes,
  grids, and ribbons produced by flow coating, Advanced Materials 22~(41)
  (2010) 4600--4604.

\bibitem{choudhary2019controlled}
S.~Choudhary, A.~J. Crosby, Controlled processing of polymer nanoribbons:
  Enabling microhelix transformations, Journal of Polymer Science Part B:
  Polymer Physics 57~(18) (2019) 1270--1278.

\bibitem{paeng2011molecular}
K.~Paeng, M.~Ediger, Molecular motion in free-standing thin films of poly
  (methyl methacrylate), poly (4-tert-butylstyrene), poly
  ($\alpha$-methylstyrene), and poly (2-vinylpyridine), Macromolecules 44~(17)
  (2011) 7034--7042.

\bibitem{bansal2005quantitative}
A.~Bansal, H.~Yang, C.~Li, K.~Cho, B.~C. Benicewicz, S.~K. Kumar, L.~S.
  Schadler, Quantitative equivalence between polymer nanocomposites and thin
  polymer films, Nature materials 4~(9) (2005) 693--698.

\bibitem{lee2013macroscopic}
D.~Y. Lee, J.~T. Pham, J.~Lawrence, C.~H. Lee, C.~Parkos, T.~Emrick, A.~J.
  Crosby, Macroscopic nanoparticle ribbons and fabrics, Advanced materials
  25~(9) (2013) 1248--1253.

\bibitem{gloumakoff1964}
N.~A. Gloumakoff, Y.-Y. Yu, {Torsion of Bars With Isosceles Triangular and
  Diamond Sections}, Journal of Applied Mechanics 31~(2) (1964) 332--334.

\bibitem{gilmour1979elastic}
I.~Gilmour, A.~Trainor, R.~Haward, Elastic moduli of glassy polymers at low
  strains, Journal of Applied Polymer Science 23~(10) (1979) 3129--3138.

\bibitem{stafford2006elastic}
C.~M. Stafford, B.~D. Vogt, C.~Harrison, D.~Julthongpiput, R.~Huang, Elastic
  moduli of ultrathin amorphous polymer films, Macromolecules 39~(15) (2006)
  5095--5099.

\bibitem{torres2009elastic}
J.~M. Torres, C.~M. Stafford, B.~D. Vogt, Elastic modulus of amorphous polymer
  thin films: relationship to the glass transition temperature, Acs Nano 3~(9)
  (2009) 2677--2685.

\bibitem{bay2020mechanical}
R.~K. Bay, K.~Zarybnicka, J.~Jancar, A.~J. Crosby, Mechanical properties of
  ultrathin polymer nanocomposites, ACS Applied Polymer Materials 2~(6) (2020)
  2220--2227.

\bibitem{kim2005deformation}
M.~Kim, T.~R. Powers, Deformation of a helical filament by flow and electric or
  magnetic fields, Physical Review E 71~(2) (2005) 021914.

\end{thebibliography}

\end{document}